\documentstyle[aps,prl,multicol,epsfig,amssymb]{revtex}

\newcommand{\PRB}[3]{Phys. Rev. B {\hspace{0.5em}\bf #1},\hspace{0.5em}#2\hspace{0.5em}({#3})}

\newcommand{\PRL}[3]{Phys. Rev. Lett. {\hspace{0.5em}\bf #1},\hspace{0.5em}#2\hspace{0.5em}({#3})}

\newcommand{\WRM}[3]{Waves Random Media {\hspace{0.5em}\bf #1},\hspace{0.5em}#2\hspace{0.5em}({#3})}

\newcommand{\ms}{{m^{\!*}}}

\newcommand{\tq}{{\tau_{\rm q}}}
\newcommand{\td}{{\tau_{\rm D}}}
\newcommand{\cmob}{{{\rm cm}^2\!/\rm Vs}}
\newcommand{\dens}{{\times10^{11} cm^{-2}}}
\newcommand{\kb}{{k_{\rm B}}}
\newcommand{\kf}{{k_{\rm F}}}
\newcommand{\ld}{{l_{\rm D}}}
\newcommand{\Ef}{{E_{\rm F}}}

\newcommand{\rxx}{{\rho_{xx}}}

\begin{document}
\title{Analysis of the resistance in p-SiGe over a wide temperature range}
\author{V. Senz$^1$, T. Ihn$^1$, T. Heinzel$^1$,
K. Ensslin$^1$, G. Dehlinger$^2$, D. Gr\"utzmacher$^2$, U. Gennser$^2$, \\E.H. Hwang$^3$, and S. Das Sarma$^3$}
\address{$^1$Solid State Physics Laboratory, ETH Z\"urich, CH-8093 Z\"urich, Switzerland\\}
\address{$^2$Paul Scherrer Institute, CH-5234 Villigen PSI, Switzerland\\}
\address{$^3$ Department of Physics, University of Maryland,  Maryland 20742-4111, USA} 
\date{\today}
\maketitle
\begin{abstract}
The temperature dependence of a system exhibiting a `metal-insulator transition in two dimensions at zero magnetic field' (MIT) is studied up to 90K. Using a classical scattering model we are able to simulate the non-monotonic temperature dependence of the resistivity in the metallic high density regime. We show that the temperature dependence arises from a complex interplay of metallic and insulating  contributions contained in the calculation of the scattering rate $1/\td(E,T)$, each dominating in a limited temperature range.\\
\end{abstract}
\begin{multicols}{2}
To date the microscopic origin of the metallic phase of what  has become known as the
`metal-insulator transition in two dimensions at zero magnetic field'
(MIT)  is still under discussion  \cite{review}. We have recently analyzed 
the temperature dependence of the metallic phase in p-SiGe \cite{Senz00} in 
terms of temperature dependent screening \cite{Gold86,Sarma99} and quantum corrections to the conductivity \cite{Aleiner98}. At low 
temperatures the temperature dependence of the conductivity could be 
quantitatively described by three contributions, namely 
$\sigma(T) = \sigma_{D}(T) + \delta\sigma_{WL}(T) + 
\delta\sigma_{I}(T),$
where $\sigma_{D}(T)$ is the Drude conductivity, 
$\delta\sigma_{WL}(T)$ and $\delta\sigma_{I}(T)$ are the weak localisation (WL) and 
the interaction contributions, respectively. It was shown that the metallic behavior in the SiGe-systems stems from screening effects contributing to $\sigma_D(T)$.
In the present study we focus on the classical Drude part of the conductivity and extend the experimental range of 
temperatures up to 90 K.
We aim at reproducing the non-monotonic temperature dependence of the experimental curves with a classical scattering description of the resistivity \cite{Sarma99}.
%In the model we consider various scattering mechanisms, such as 
%ionized impurity scattering, interface
%roughness scattering, alloy disorder scattering, ionized background
%impurity scattering and acoustic phonon scattering.
 In this way we 
are able to make a quantitative comparison with the experimental 
data in a wide range of temperatures ($1.7 K \le T \le 
20 K$) and densities ($1.5 \times 10^{11} cm^{-2} \le p\le 5.1 
\times 10^{11} cm^{-2}$).
 Below 4K temperature dependent 
screening is the dominant contribution to the temperature 
dependence of the resistivity. At higher temperatures the smearing of the
Fermi function and phonon scattering produce a 
non-monotonic temperature dependence.
 The good agreement of theory with experiment indicates that temperature dependent screening cannot 
only account for the metallic behavior close to the 
metal-insulator transition but is equally important in adjacent 
temperature and density regimes.
The model cannot be applied  in the insulating regime
($\kf \ld<1$, where $\kf$ is Fermi wave vector and $\ld$ Drude mean free path).

\vspace{1ex}
In our MBE-grown samples the 2DHG resides in a 20nm Si$_{0.85}$Ge$_{0.15}$ quantum well sandwiched between two undoped Si layers. Remote Boron doping was introduced at a distance of 15nm above the quantum well and a Ti/Al Schottky gate allowed tuning the hole density.
For the transport measurements standard Hall-bars were used (L=300$\mu$m, W=300$\mu$m). The experiments were carried out in a $^4$He-system in a temperature range of 1.7K - 90K, using standard four terminal AC-technique.
The hole density in the 2DHG was adjustable from complete depletion to a density of $5.1\dens$, yielding a maximum mobility of $7800\cmob$ at the highest density. The ratio of the Drude scattering time $\td$ and Shubnikov-de Haas relaxation time $\tq$ is close to one, indicating that large angle scattering dominates the transport in such structures \cite{emeleus,whall}. The insulating contributions of the weak localization effect and electron-electron interaction become significant at temperatures $<$ 1K. For a detailed study of these effects see Ref. \cite{Senz00}.
%{\footnotesize\begin{picleft}[0.5\mpbreite]{B509L17KTsw.eps}{Temperature sweeps for different densities $p=5.1,4.2,3.3,2.7,2.4,2.3,2.1,2.0\dens$; the long range sweep (90K-1.7K, thin line) is done at B=0.5T, the short range sweep (10K-1.7K, thick line) at B=0T. The corresponding Fermi temperature $\Tf$, taken at the lowest temperature, for each trace is marked with a blue filled circle, the Dingle temperature $T_D=\Gamma/\kb$ with $\Gamma=\hbar/\tq$ is marked with a filled black circle.\label{fig1}}\end{picleft}}
\begin{figure}[h!]
\includegraphics[width=0.9\linewidth]{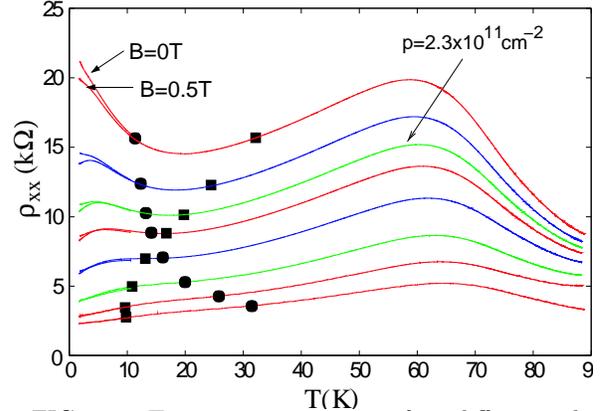}
\caption{Temperature sweeps for different densities $p=5.1-2.0\dens$; the long range sweep (90K-1.7K, thin line) is done at B=0.5T, the short range sweep (10K-1.7K, thick line) at B=0T.  The WL-contribution acts localizing and is suppressed at B=0.5T. {\Large$\bullet$} represent the temperature $T_c$ above which the system becomes non-degenerate, $\blacksquare$ marks the Dingle temperature $T_D$. \label{fig1}}
\end{figure}

\begin{figure}[h!]
\includegraphics[width=0.9\linewidth]{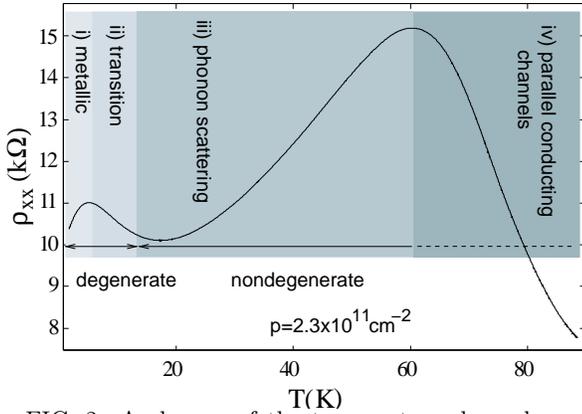}
\caption{A closeup of the temperature dependence of the resistivity for the density $p=2.3\dens$. Metallic and insulating behavior alternates depending on the temperature range.\label{fig2}}
\end{figure}

\vspace{1ex}
Fig. \ref{fig1} shows measurements of the longitudinal resistance $\rxx$ as a function of temperature between 1.7K and 90K for different densities $p$.
A typical curve (e.g. $p=2.3\dens$, see  Fig. \ref{fig2}) can be divided into four different temperature ranges:
\renewcommand{\theenumi}{\roman{enumi}}
\begin{enumerate}
\item T$<$4K: the temperature dependence is metallic.
\item 4K$<$T$<$14K: the temperature dependence is non-monotonic and a turnover to insulating behavior at higher temperatures occurs.
\item 14K$<$T$<$60K: the dependence changes again to metallic behavior.
\item 60K$<$T$<$90K: in this regime the 2DHG is not the only conducting channel in the sample (visible in the Hall density, not shown). This leads to a strong resistivity decrease as the temperature is raised from 60K to 90K .
\end{enumerate}

\vspace{1ex}
In order to achieve a detailed understanding of these results  we apply the standard semi-classical scattering model for the resistivity \cite{Sarma99,ando} which is valid in the limit of $\kf\ld>1$.\\
For finite temperatures and zero magnetic field the resistivity is expressed as
\[\rho_{xx}(T) = \frac{\ms}{pe^2\left<\tau(T)\right>},\]
with the effective mass of $\ms=0.25m_0$ (obtained from the T-dependence of Shubnikov-de Haas oscillations) and the scattering time
\begin{equation} 
\left<\tau(T)\right> = \frac{1}{E_{F}}\int_{0}^\infty dE E 
\tau(E,T)\left(-\frac{df(E,\mu,T)}{dE}\right),
\label{equ1}
\end{equation}
where $f(E,\mu,T)=(e^{(E-\mu)/\kb T} +1)^{-1}$ is the Fermi distribution function ($\mu$ temperature dependent chemical potential, $\kb$ Boltzmann constant) and $E_F$ is the Fermi energy.\\
The transport scattering rate is calculated according to
\begin{equation}
\frac{1}{\tau(E,T)} = \frac{\ms}{\pi\hbar^3}\int_{0}^{\pi}d\theta 
\frac{\left<\left|V(q_E)\right|^2\right>}{\varepsilon^2(q_E,T)}(1-\cos\theta),
\label{equ2}
\end{equation}
where $q = \sqrt{4\ms E(1-\cos\theta)}/\hbar$ is the momentum transfer in a scattering event.\\
In the calculation we consider ionized impurity scattering \cite{Gold86}, interface roughness scattering \cite{Gold86}, alloy disorder scattering \cite{ando1}, ionized background impurity scattering \cite{Gold87} and acoustic phonon scattering \cite{Kamamura}. They enter into $<\left|V(q_E)\right|^2>$ which represents the sum of all individual mechanisms.

In this formulation the T-dependence of $\rxx$ arises due to the energy averaging in Eq. \ref{equ1} and due to the explicit T-dependence of Lindhard's dielectric function
\begin{equation}\epsilon(q,T) =1+V(q)\Pi(q,T)F(q)\left [1-G(q) \right ],
\label{epsequ}
\end{equation}
where $V(q)=2\pi e^2/\varepsilon q$ is the interaction potential of the hole gas in two dimensions.
The T-dependence in Eq. \ref{epsequ} again is due to energy averaging when calculating the polarizability function \cite{maldague}
\begin{equation}\Pi(q,T) =\int_{0}^\infty 
dE\;\Pi(q,T=0,E) \left(-\frac{df(E,\mu,T)}{dE}\right) .
\label{piequ}\end{equation}
In Eq. \ref{epsequ} 
$F(q)$ is the form factor for the inversion layer\cite{ando},
%We use the variational wave function for the inversion layer \cite{ando}
%$\zeta_{0}(z) = \sqrt{\frac{b^3}{2}}ze^{-bz/2}$  for our model \cite{ando}, 
%where the width parameter is given by 
%$b =[(12\pi \ms e^2)/(\pi\hbar^2\varepsilon\varepsilon_{0})\cdot(N_{depl}
%+11n_s/32)]^{1/3}$ ($N_{depl}$ is the number of charges per unit area in 
%the depletion layer).
% The form factor then becomes $F(q) = (8+9w+3w^2)/8(1+w)^3$ with $w=q/b$. 
and $G(q) = q/g\sqrt{q^2+k_{F}^2}$ with 
a degeneracy factor $g$ is 
the local field correction in the Hubbard-approximation \cite{Gold86} .
\begin{figure}[h!]
\includegraphics[width=0.8\linewidth]{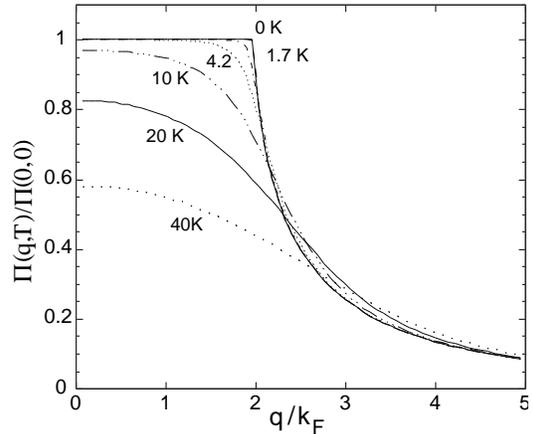}
\caption{Polarizability function $\Pi(q,T)$ for various temperatures  
and the density $p=2.3\times 10^{11} cm^{-2}$.\label{fig3}}
\end{figure}

\vspace{1ex}
Let us look in more detail into the Eqs. \ref{equ1} and \ref{piequ}.  We can assert two general effects for the temperature dependence: (1) the temperature dependence of the chemical potential $\mu(T)=\kb T \ln (e^{E_F/\kb T}-1)$: this shifts the balance point around which the integral is performed to lower energies with increasing temperature (2)  the integration itself performs an energy averaging around $\mu(T)$ with width of about $4\kb T$ and depends strongly on the curvature of the integrand. In both cases the structure is the same: the integrand is a product of the symmetric (around $E=\mu$) function $df/dE$  with an asymmetric function, i.e. $E\tau$ and $\Pi$, which in total yields asymmetric functions but with opposite sign in the curvature\\
 The curvature of $\Pi(q,T)$ is negative (see  Fig. \ref{fig3}) meaning that  $d\Pi/dT<0\Rightarrow d\epsilon/dT<0\Rightarrow d\tau/dT<0$ due to energy averaging. The temperature dependence of the chemical potential  leads  to a slight damping of the energy averaging effect  but can not change its overall behavior. In our case it is negligible:  the effect becomes only important for temperatures higher than $T_c$ at which the chemical potential $\mu(T)$ deviates about 10\% from its zero temperature value $\mu(0)=\Ef$ (indicated by filled circles in  Fig. \ref{fig1}).
\begin{figure}[h!]
\includegraphics[width=0.8\linewidth]{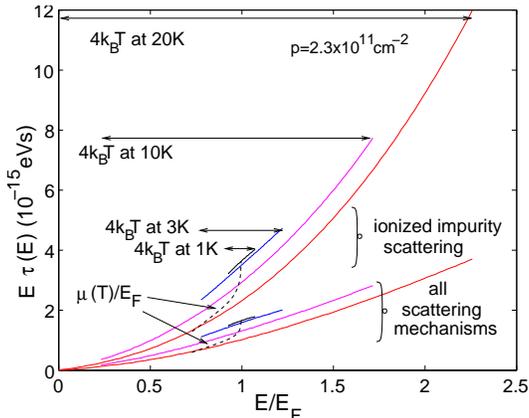}
\caption{The influence of different scattering contributions $V(q)$ on the integrand of Eq. \ref{equ1}, $E\tau_e(E,T)$, for different temperatures.\label{fig4}}
\end{figure}
 Due to reduced screening  $\tau(E,T)$ in Eq. \ref{equ1} decreases with increasing temperature, but energy averaging counteracts this tendency since  the curvature of $E\tau(E,T)$ is positive and  therefore $d\left<\tau\right>/dT>0$. We illustrate this behavior of $E\tau(E,T)$ in  Fig. \ref{fig4}. At the lowest temperatures  the temperature dependence of the polarizability function leads to a metallic behavior in the resistivity.  In the special case of large angle scattering, i.e. $q=2\kf$ which is dominant in SiGe samples \cite{Gold86}, the temperature dependence of $ \Pi(q,T)$ is strongest. Only at higher temperatures the insulating contribution of the energy averaging becomes significant and finally overcomes the metallic screening effect.

\vspace{1ex}
We now discuss the interplay of these effects in the temperature ranges (i)-(iii) mentioned in the beginning:
\begin{enumerate}
\item 2DHG is degenerate: the screening function dominates the temperature behavior. For  large angle scattering (i.e. scattering for $q\approx 
2k_{F}$) it has been shown that \cite{Senz00,Gold87} 

$\sigma_{D}(T) = \sigma_{D}(0)\left[1-C(p)\frac{T}{T_{F}}\right]+{\cal O}\left[(\frac{T}{T_{F}})^{3/2}\right].$ 
\vspace{1ex}
\item  the effect of energy averaging comes into play and counteracts temperature dependent screening. At higher temperatures energy averaging dominates, i.e. the temperature dependence is insulating.
\item the system becomes non-degenerate: phonon scattering is the strongest scattering mechanism; the observed temperature dependence is due to phonon freeze out. This behavior is in agreement with the generally accepted linear temperature dependence of acoustic deformation potential scattering for which a larger temperature coefficient of the resistivity is expected for lower carrier densities \cite{ando}.
\end{enumerate}
Deviations from this scenario occur at the two extreme densities. At high density the maximum in the resistivity disappears gradually and a purely metallic temperature dependence is observed over the entire T-range. At the lowest densities one enters the strongly localized regime where the presented model can no longer be applied.

\begin{figure}[h!]
\includegraphics[width=0.8\linewidth]{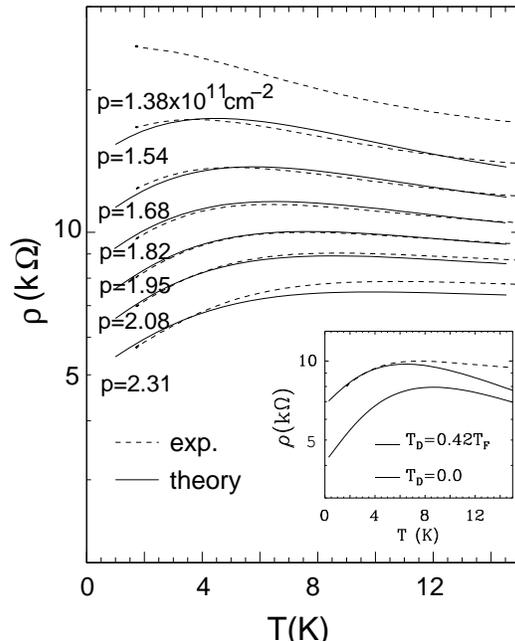}
\caption{Comparison between experiment and calculations. 
In this calculations we include interface charge and phonon scattering.
We consider the collisional broadening effect on screening with
$T_D=0.8,\;0.6,\;0.52,\;0.42,\;0.37,\;0.34T_F$ (from top to bottom). 
In inset the effect of level broadening 
is shown for the density $p=1.95\dens$. Here only interface 
charge scattering is considered.
\label{fig5}}
\end{figure}
We are now able to compare our calculations with the experiment. This is illustrated in  Fig. \ref{fig5}. In order to achieve quantitative agreement between experiment and calculations we additionally included level-broadening effects on the polarizability function characterized by the parameter $\Gamma=\hbar/\tq=\kb T_D$ ($T_D$ Dingle temperature, marked with filled squares in  Fig. \ref{fig1}) \cite{Sarma86}. This acts similar to thermal effects in rounding off the sharp corner of the polarizability function visible in  Fig. \ref{fig3}. The influence on the resistivity can be seen in the inset in  Fig. \ref{fig5}: it reduces the effect of temperature dependent screening and the overall resistance rises.

The agreement of our simulations with the experiment is, even quantitatively, quite good. Only at the lowest densities the model fails in describing the experimental data, due to the reasons mentioned above.

\vspace{1ex}
In summary we have shown that the non-monotonicity in the temperature dependence of the resistivity of p-type SiGe samples showing a MIT can be described by the temperature dependence of the scattering time $\td$. 
The metallic part stems from the energy averaging of the polarizability function $\Pi(q,T)$, the insulating part from the energy averaging when performing the integral  for $\tau(E,T)$. This interplay is not significantly altered by the T-dependence of the chemical potential $\mu(T)$. Finally we compare our calculations with experimental curves and find quantitatively good agreement. 

Financial support from ETH Z\"urich and the Schweizerische Nationalfonds is gratefully acknowledged.

%{\footnotesize\begin{pictwo}[0.5\mpbreite]{B509L17KTsw.eps}{Temperature sweeps for different densities $p=5.1,4.2,3.3,2.7,2.4,2.3,2.1,2.0\dens$; the long range sweep (90K-1.7K, thin line) is done at B=0.5T, the short range sweep (10K-1.7K, thick line) at B=0T.  {\Large$\bullet$} represent the temperature $T_c$ above which the system becomes non-degenerate, $\blacksquare$ marks the Dingle temperature $T_D$ \label{fig1}}{Tswe17KpaperMSSinsetmod.eps}{A closeup of the temperature dependence of the resistivity for the density $p=2.3\dens$. Metallic and insulating behavior alternates depending on the temperature range.\label{fig2}}\end{pictwo}}
%{\footnotesize\begin{pictwo}[0.5\mpbreite]{polfun.eps}{Polarization function $\Pi(q,T)$ for various temperatures.\label{fig3}}{Scaateringtimespapermssmod.eps}{The influence of different scattering contributions $V(q)$ on the integrand of Eq. \ref{equ1}, $E\tau_e(E,T)$, for different temperatures.\label{fig4}}\end{pictwo}}
%{\footnotesize\begin{picleft}[0.4\mpbreite]{fig5mod.eps}{Comparison between experiment and calculations. For the calculations only the main scattering contributions interface charge and phonon scattering were included. In the inset the effect of level broadening  represented by $T_D$ is shown for the density $p=1.95\dens$.\label{fig5}}\end{picleft}}
\end{multicols}
\end{document}